\documentclass[prl,aps,floats,twocolumn]{revtex4}
\usepackage{amssymb}
\usepackage{graphicx}
\usepackage{amsmath}
\usepackage{amsfonts}
\usepackage{accents}

\begin{document}
%\title{Fundamental limits of quantum communication \\and general benchmarks for quantum repeaters}
\title{General Benchmarks for Quantum Repeaters}
\author{Stefano Pirandola}
\author{Riccardo Laurenza}
\affiliation{Computer Science and York Centre for Quantum Technologies, University of York,
York YO10 5GH, United Kingdom}

\begin{abstract}
Using a technique based on quantum teleportation, we simplify the most general
adaptive protocols for key distribution, entanglement distillation and quantum
communication over a wide class of quantum channels in arbitrary dimension.
Thanks to this method, we bound the ultimate rates for secret key generation
and quantum communication through single-mode Gaussian channels and several
discrete-variable channels. In particular, we derive exact formulas for the
two-way assisted capacities of the bosonic quantum-limited amplifier and the
dephasing channel in arbitrary dimension, as well as the secret key capacity
of the qubit erasure channel. Our results establish the limits of quantum
communication with arbitrary systems and set the most general and precise
benchmarks for testing quantum repeaters in both discrete- and
continuous-variable settings.

\end{abstract}
\maketitle

%\title{General benchmarks for quantum repeaters in bosonic and qubit channels}
%\title{Funbdamental benchmarks for quantum repeaters}
%\title{Limits of quantum communication and benchmarks for quantum repeaters}

%\title{Teleportation stretching of quantum communication}

%\subsection{Introduction}

Today's quantum technologies are the subject of an increasing interest by the
scientific community, and represent one the biggest investments of research
funding bodies. The promise of reliable and completely secure quantum
communications~\cite{BB84,Ekert,Fred,MDI1,CVMDIQKD} is an appealing
alternative to the present classical infrastructure. The idea of developing a
quantum Internet~\cite{Kimble2008}, which might lead to a world-wide
architecture for distributed quantum processing, is attracting huge efforts
from different fields, such as quantum optics and condensed matter physics.
Hybrid solutions based on different substrates and technologies are considered
the best strategies in this
direction~\cite{telereview,Ulrikreview,XiangRMP13,KurizkiPNAS15}.

However, the problem is that quantum information is generally more fragile
than its classical counterpart. Exotic quantum features may be rapidly washed
away by the inevitable interactions with the external environment. Noise and
decoherence may greatly limit the performances of the various quantum
protocols, as quantified by their optimal rates for transmitting quantum
information, entanglement or secret correlations. This is an underlying
limitation which potentially affect any direct quantum communication between
two parties, say Alice and Bob. For this reason one needs to consider quantum
repeaters~\cite{Briegel98}, i.e., intermediate relay stations which are meant
to assist any point-to-point implementation.

This work is directly addressed to the core of this problem. We investigate
the optimal rates which are achievable for quantum communication (QC),
entanglement distillation\ (ED) or quantum key distribution (QKD) between two
parties, who can exploit unlimited two-way classical communication (CC) and
adaptive local operations (LOs), also known as adaptive LOCCs. We assume that
the two parties are directly connected by a quantum channel and they do not
pre-share any entanglement. In such a situation, we study the various two-way
assisted capacities of the channel for QC, ED and QKD which set the ultimate
limits achievable in the absence of repeaters.

By applying a technique of teleportation stretching~\cite{QKDpaper} to
suitable \textquotedblleft stretchable\textquotedblright\ channels in
arbitrary dimension, we are able to greatly reduce the complexity of any
quantum protocol based on adaptive LOCCs. The result is a simple protocol
where each transmission through the channel is replaced by a Choi matrix, and
the adaptive LOCCs are all collapsed into a single final LOCC. This allows us
to bound the optimal rates for QC, ED and QKD by a single quantity, that we
call \textquotedblleft entanglement flux\textquotedblright\ and represents the
maximum entanglement that may survive the channel. This quantity is computed
for an arbitrary single-mode Gaussian channel (extending the results of
Ref.~\cite{QKDpaper}) and for several discrete-variable channels, including
all qubit Pauli channels.

By showing coincidence with lower bounds, we prove a number of exact formulas.
We show that the two-way assisted quantum capacity $Q_{2}$ and the secret-key
capacity $K$\ of the quantum-limited amplifier are $Q_{2}=K=\log_{2}[g/(g-1)]$
where $g$ is the gain. We then show that the qubit dephasing channel has
two-way assisted capacities $Q_{2}=K=1-H_{2}(p)$, where $H_{2}$ is the binary
Shannon entropy and $p$ is the dephasing probability. We also extend this
result to arbitrary dimension.\ Finally, we also determine the secret-key
capacity of the qubit erasure channel, which is $K=1-p$. It is worth to
mention that, before our study, the only two-way assisted capacities that were
known were those ($K$ and $Q_{2}$) of the pure-loss channel~\cite{QKDpaper}
and $Q_{2}$ of the erasure channel~\cite{ErasureChannel}.

\subsection{Adaptive quantum communication in arbitrary dimension}

Suppose that Alice and Bob are separated by a quantum channel $\mathcal{E}$
and they want to implement the most general protocol assisted by adaptive
LOCCs, with the aim of distributing entanglement, quantum information or
secret keys. We assume that Alice and Bob have countable sets of systems,
denoted by $\mathbf{a}$ and $\mathbf{b}$, respectively.

The first step is the preparation of the initial state of $\mathbf{a}$ and
$\mathbf{b}$ by LOCC $\Lambda_{1}$. Next, Alice picks a system $a_{1}%
\in\mathbf{a}$ which is sent through the channel $\mathcal{E}$. Once Bob gets
the output $b_{1}$, the parties apply LOCC $\Lambda_{2}$ on all systems
$\mathbf{a}b_{1}\mathbf{b}$. Let us update Bob's set by including $b_{1}$,
i.e., $b_{1}\mathbf{b}\rightarrow\mathbf{b}$. In the second transmission,
Alice sends another system $a_{2}\in\mathbf{a}$ through $\mathcal{E}$
resulting into an output $b_{2}$ for Bob. The parties apply further LOCC
$\Lambda_{3}$\ on all systems $\mathbf{a}b_{2}\mathbf{b}$. Bob's set is
updated and so on. After $n$ transmissions, Alice and Bob will share a state
$\rho_{\mathbf{ab}}^{n}$ depending on the sequence of adaptive LOCCs
$\mathcal{L}=\{\Lambda_{1},\cdots,\Lambda_{n+1}\}$.

This adaptive protocol has a rate of $R^{n}$ if $\left\Vert \rho_{\mathbf{ab}%
}^{n}-\phi_{n}\right\Vert \leq\varepsilon$, where $\left\Vert \cdot\right\Vert
$ is the trace norm and $\phi_{n}$ is a target state with $nR^{n}$ bits. If
the parties implement entanglement distillation (ED), the target state is a
maximally-entangled state and $R_{\text{ED}}^{n}$ is the number of
entanglement bits (ebit) per use. If the parties implement QKD, the target
state is a private state~\cite{KD1} with secret-key rate $R_{\text{K}}^{n}\geq
R_{\text{ED}}^{n}$~\cite{notaME}. By taking the limit of $n\rightarrow+\infty
$\ and optimizing over all the protocols $\mathcal{L}$, one can define the
two-way entanglement distillation capacity $D_{2}$ and the secret-key capacity
$K$ of the channel as follows%
\begin{equation}
D_{2}(\mathcal{E}):=\sup_{\mathcal{L}}\lim_{n}R_{\text{ED}}^{n}\leq
K(\mathcal{E}):=\sup_{\mathcal{L}}\lim_{n}R_{\text{K}}^{n}. \label{assisted}%
\end{equation}

Note that an ebit can teleport a qubit and a qubit can distribute an ebit:
These processes are fully equivalent in the presence of unlimited two-way CCs.
Thus, $D_{2}(\mathcal{E})$ coincides with the two-way quantum capacity
$Q_{2}(\mathcal{E})$ of the channel. All these capacities can be bounded by
introducing suitable quantities. From below we may consider the (reverse)
coherent information~\cite{Schu96,Lloyd97,RevCohINFO,ReverseCAP} of the
channel $I_{(R)C}(\mathcal{E})$, which we define as the (reverse) coherent
information computed on its Choi matrix (see Methods for definitions). In
particular, we have $I_{C}(\mathcal{E})\leq Q^{(1)}(\mathcal{E})$, where
$Q^{(1)}$ is the one-shot unassisted quantum capacity of the channel. These
quantities not only are computable but, thanks to the hashing
inequality~\cite{DWrates}, they represent achievable rates for one-way
entanglement distillation and, therefore, are lower bounds of $D_{2}%
(\mathcal{E})$.

From above we may resort to the relative entropy of entanglement
(REE)~\cite{RMPrelent}. Recall that, for any bipartite state $\rho$, this is
defined as $E_{R}(\rho)=\min_{\sigma\in\text{SEP}}S(\rho||\sigma)$, where SEP
are separable states and $S(\rho||\sigma):=\mathrm{Tr}\left[  \rho(\log
_{2}\rho-\log_{2}\sigma)\right]  $ is the relative entropy~\cite{WildeBOOK}.
Then, Ref.~\cite{QKDpaper} showed that
\begin{equation}
K(\mathcal{E})\leq E_{R}(\mathcal{E}):=\sup_{\mathcal{L}}\lim\sup_{n}%
n^{-1}E_{R}(\rho_{\mathbf{ab}}^{n})~, \label{hard}%
\end{equation}
in any dimension. The REE bound $E_{R}(\mathcal{E})$ is generally
very hard to compute. Remarkably, its calculation can be
enormously simplified if the channel suitably \textquotedblleft
commutes\textquotedblright\ with teleportation.

\begin{figure*}[ptbh]
\vspace{-4.2cm}
%\vspace{-0.2cm}
\par
\begin{center}
\includegraphics[width=0.98\textwidth] {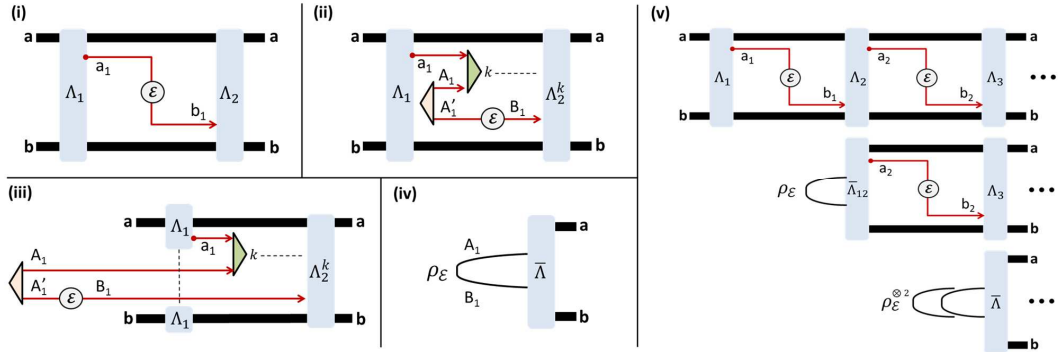}
\end{center}
\par
\vspace{-4.2cm}
%\vspace{-0.2cm}
\caption{\textbf{Stretching of an adaptive protocol. }Time flows
from left to right, Alice is at the top and Bob at the bottom. In
panel~(i) we show the first transmission $a_{1}\rightarrow b_{1}$
over channel $\mathcal{E}$, which occurs between two LOCCs
$\Lambda_{1}$ and $\Lambda_{2}$ performed by the parties on their
ensembles of systems $\mathbf{a}$ and $\mathbf{b}$. In panel~(ii)
we insert an ideal teleportation circuit, just before the channel.
This is composed by an ideal EPR state $\Phi^{\text{EPR}}$ of
systems $A_{1}$ and $A_{1}^{\prime}$ (orange triangle), and a Bell
detection performed on systems $a_{1}$ and $A_{1}$ (green
triangle). As a result, $a_{1}$ is perfectly teleported into the
new input $A_{1}^{\prime}$ up to a $k $-dependent unitary $T_{k}$.
Since $\mathcal{E}$\ is stretchable, $T_{k}$ is mapped into an
output unitary $U_{k}$ on system $B_{1}$. This unitary is erased
by Bob in the second modified LOCC $\Lambda_{2}^{k}$ upon receving
the CC of $k$ from Alice. In panel~(iii) we anticipate the
distribution of the EPR source and post-pone the Bell detection
after the channel. In panel~(iv) we show the final result, which
is a simple protocol
where the Choi-matrix of the channel $\rho_{\mathcal{E}}=(\mathcal{I}%
\otimes\mathcal{E})(\Phi^{\text{EPR}})$ is subject to a final LOCC $\Lambda$,
combining the previous adaptive LOCCs. The action of $\Lambda$ over
$\rho_{\mathcal{E}}$ does not depend on the outcome $k$. This means that we
can replace $\Lambda$\ by the mean LOCC $\bar{\Lambda}$ averaged over the Bell
outcomes, which is trace-preserving~\cite{NotaCOH,Differenza}. The full
stretching of the protocol can be done iteratively. As shown in panel (v),
once the first transmission is stretched, it becomes the input for the second
transmission, which is in turn stretched into another Choi-matrix. All the
adaptive LOCCs $\Lambda_{1}$, $\Lambda_{2}$ and $\Lambda_{3}$, reduce to a
single final trace-preserving LOCC $\bar{\Lambda}$, applied to all
Choi-matrices. The extension to arbitrary $n$ is straightforward and leads to
the output state in Eq.~(\ref{outLEMMA}).}%
\label{panpic}%
\end{figure*}

\subsection{Stretching of adaptive protocols}

Denote by $\mathcal{S}$\ the set of teleportation unitaries in arbitrary
dimension $d$. For a qudit ($d<+\infty$), the teleportation set $\mathcal{S}$
is composed by generalized Pauli operators $T_{k}$ with $k=1,\ldots,2^{d}$.
These are the generators of a finite-dimensional Weyl-Heisenberg group or
Pauli group (see Methods). For a bosonic system ($d=+\infty$), the set
$\mathcal{S}$ is composed by displacement operators $T_{k}$ with complex
$k$~\cite{telereview}, which form the infinite-dimensional Weyl-Heisenberg
group. \ By definition, we say that a quantum channel $\mathcal{E}$ is
\textquotedblleft stretchable\textquotedblright\ by teleportation\ if, for any
$T_{k}\in\mathcal{S}$ and any input state $\rho$, we may write%
\begin{equation}
\mathcal{E}(T_{k}\rho T_{k}^{\dagger})=U_{k}\mathcal{E}(\rho)U_{k}^{\dagger}~,
\label{stretch}%
\end{equation}
for some unitary $U_{k}$.

Typically, the condition of Eq.~(\ref{stretch}) is satisfied with $U_{k}%
\in\mathcal{S}$, i.e., the channel is covariant with the Weyl-Heisenberg
group. All qubit Pauli channels are stretchable, because Pauli operators
commute or anticommute one with each other. Then, all bosonic Gaussian
channels are stretchable since they linearly transform the quadratures, so
that input displacements are mapped into output ones. Clearly, there exist
channels which are not stretchable, an example being the amplitude damping
channel~\cite{AmplitudeDamp}.

Quantum communication over a stretchable channel can be greatly
simplified. In fact, the previous adaptive protocol can be
suitably \textquotedblleft stretched\textquotedblright\ in time
and reduced into a non-adaptive protocol where channels are
replaced by their Choi matrices and the adaptive LOCCs are
collapsed into a single final LOCC. This is possible by
introducing, before each use of the channel, an ideal
teleportation circuit which is composed by an ideal EPR source,
i.e., a maximally-entangled state, and a corresponding Bell
detection. The whole process is shown in Fig.~\ref{panpic}. See
Methods for more details on ideal teleportation and the detailed
maths of teleportation stretching.

Remarkably, an adaptive protocol over $n$ uses of a stretchable channel
$\mathcal{E}$ reduces to $n$ Choi matrices $\rho_{\mathcal{E}}$ plus a
trace-preserving LOCC $\bar{\Lambda}$, i.e., the output reads%
\begin{equation}
\rho_{\mathbf{ab}}^{n}=\bar{\Lambda}\left(  \rho_{\mathcal{E}}^{\otimes
n}\right)  . \label{outLEMMA}%
\end{equation}
This reduction of the output state greatly simplifies the
computation of any entanglement measure which is non-increasing
under trace-preserving LOCC and subadditive on tensor products.
This is exactly the case of the REE, for which we may write
$E_{R}(\rho_{\mathbf{ab}}^{n})\leq nE_{R}(\rho_{\mathcal{E}})$. By
replacing in Eq.~(\ref{hard}), we obtain $K(\mathcal{E})\leq E_{R}%
(\mathcal{E})=E_{R}(\rho_{\mathcal{E}})$.

Now, let us define the entanglement flux $\Phi(\mathcal{E})$ of a channel
$\mathcal{E}$ as the REE of its Choi matrix, i.e., $\Phi(\mathcal{E}%
):=E_{R}\left(  \rho_{\mathcal{E}}\right)  $. For a stretchable
channel, this quantity represents the maximum amount of
entanglement (as quantified by the REE)\ which can be distributed
through the channel by means of adaptive protocols, i.e.,
$E_{R}(\mathcal{E})$. It is clearly zero for an
entanglement-breaking channel, while it is maximum for the
identity channel $\mathcal{I}$, for which
$\Phi(\mathcal{I})=\log_{2}d$ with $d$ being the dimension of the
Hilbert space (unbounded for CVs). By combining all previous
results, we have that the two-way assisted capacities of a
stretchable channel $\mathcal{E}$ are upperbounded by its
entanglement flux, i.e.,
\begin{equation}
\max\{I_{C},I_{RC}\}\leq D_{2}=Q_{2}\leq K\leq\Phi. \label{Eqtheo}%
\end{equation}

Note that the entanglement flux is convex with respect to compositions of
quantum channels. In other words, for any mean channel $\overline{\mathcal{E}%
}=\sum_{i}p_{i}\mathcal{E}_{i}$, defined for an ensemble of channels
$\{p_{i},\mathcal{E}_{i}\}$, we may write%
\begin{equation}
\Phi(\overline{\mathcal{E}})\leq\sum_{i}p_{i}\Phi(\mathcal{E}_{i}).
\label{convex}%
\end{equation}
In particular, for ensembles of stretchable channels, the mean channel
$\overline{\mathcal{E}}$ is also stretchable. Thus, combining
Eqs.~(\ref{Eqtheo}) and~(\ref{convex}), one has an upper bound for the two-way
assisted capacities of $\overline{\mathcal{E}}$.

\subsection{Bosonic Gaussian channels}

We can now investigate the ultimate rates of quantum communication and secret
key generation over the most important channels. In particular, we consider
here all single-mode Gaussian channels in canonical form~\cite{RMP,HolevoBOOK}%
, extending the analysis of Ref.~\cite{QKDpaper}. We present the results for
the most important forms (amplifier and additive-noise channels) leaving
secondary forms in the Methods.

For bosonic systems, the ideal EPR state $\Phi^{\text{EPR}}$ has infinite
energy, which means that the computation of the entanglement flux of a
Gaussian channel involves an asymptotic limit. This is done by considering a
sequence of finite-energy EPR states $\Phi_{\mu}^{\text{EPR}}$, which are
two-mode squeezed vacuum states with variance $\mu$~\cite{RMP}. In the limit
of large $\mu$, this sequence converges to $\Phi^{\text{EPR}}$ in trace norm,
so that we can exploit the lower semicontinuity of the relative entropy and
write $\Phi(\mathcal{E})=E_{R}\left(  \rho_{\mathcal{E}}\right)  \leq\lim
\inf_{\mu}E_{R}\left(  \rho_{\mu}\right)  $ where $\rho_{\mu}:=(\mathcal{I}%
\otimes\mathcal{E})(\Phi_{\mu}^{\text{EPR}})$.

In this way, Ref.~\cite{QKDpaper} determined a weak converse rate for quantum
communication over a thermal-loss channel $\mathcal{E}_{\text{loss}}$ with
transmissivity $0\leq g\leq1$\ and thermal noise $\bar{n}$. More precisely,
Ref.~\cite{QKDpaper} found the upper bound%
\begin{equation}
\Phi(g,\bar{n})\leq-\log_{2}\left[  (1-g)g^{\bar{n}}\right]  -h(\bar
{n})\text{~~for~}\bar{n}<\frac{g}{1-g}, \label{LossUB}%
\end{equation}
and $\Phi(g,\bar{n})=0$ otherwise. In the previous formula we set
$h(x):=(x+1)\log_{2}(x+1)-x\log_{2}x$. For the pure-loss channel ($\bar{n}%
=0$), this bound coincides with the reverse coherent information of the
channel, therefore establishing all its two-way capacities $Q_{2}%
(g)=K(g)=-\log_{2}(1-g)$.

Note that free-space satellite communications in the presence of atmospheric
turbulence may be modelled as an average of pure-loss channels $\mathcal{E}%
_{i}$ with transmissivities $g_{i}\in\lbrack0,1]$ and associated probabilities
$p_{i}$~\cite{Satellite}. For such a channel $\overline{\mathcal{E}}=\sum
_{i}p_{i}\mathcal{E}_{i}$ we can exploit the convexity of the entanglement
flux and write the upper bound
\begin{equation}
\Phi(\overline{\mathcal{E}})\leq-\sum_{i}p_{i}\log_{2}(1-g_{i})~.
\end{equation}

Consider the amplifier channel $\mathcal{E}_{\text{amp}}$, whose action on
input quadratures $\mathbf{\hat{x}}\rightarrow\sqrt{g}\mathbf{\hat{x}}%
+\sqrt{g-1}\mathbf{\hat{x}}_{E}$ where $g>1$ is the gain and $E$ is a thermal
environment with $\bar{n}$ mean photons. For this channel, we compute%
\begin{equation}
\Phi(g,\bar{n})\leq\log_{2}\left(  \dfrac{g^{\bar{n}+1}}{g-1}\right)
-h(\bar{n})\text{~~for~}\bar{n}<(g-1)^{-1}\text{,} \label{ampliBOUND}%
\end{equation}
and $\Phi(g,\bar{n})=0$\ otherwise (see Methods). The best known lower bound
is given by the coherent information of the channel $I_{C}(\mathcal{E}%
_{\text{amp}})=\log_{2}[g/(g-1)]-h(\bar{n})$~\cite{HolevoWerner}.

In particular, for the quantum-limited amplifier ($\bar{n}=0$), we find that
the previous upper and lower bounds coincide, thus determining its two-way
assisted capacities%
\begin{equation}
Q_{2}(g)=K(g)=\log_{2}\left(  \dfrac{g}{g-1}\right)  . \label{ampliMAIN}%
\end{equation}
They turn out to coincide with the unassisted quantum capacity $Q$ of the
channel~\cite{HolevoWerner,Wolf}. The result of Eq.~(\ref{ampliMAIN}) sets the
fundamental limit for secret-key generation, entanglement distillation and
quantum communication with amplifiers. A trivial consequence of the formula is
that infinite amplification is useless. For an amplifier with typical gain
$2$, the maximum achievable rate for quantum communication is just $1$ qubit
per use.

\begin{figure}[ptbh]
\vspace{-0.0cm}
\par
\begin{center}
\includegraphics[width=0.43\textwidth] {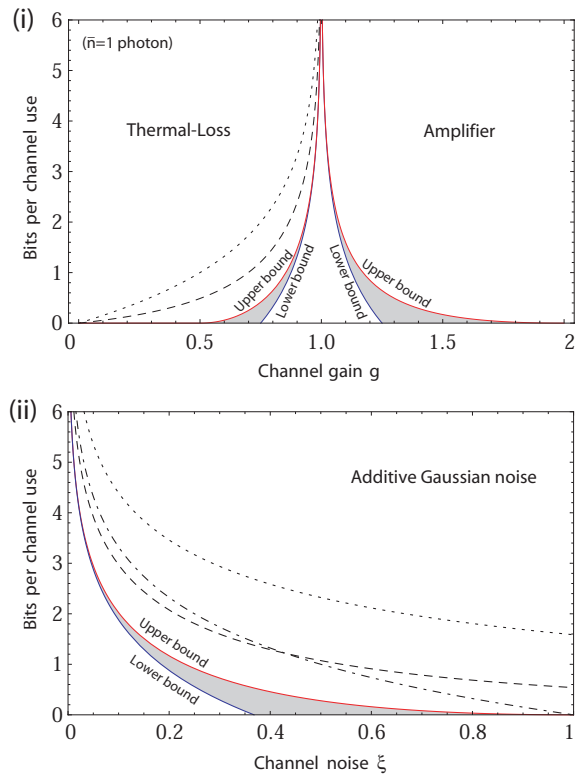}
\end{center}
\par
\vspace{-0.5cm}\caption{Plots of the bounds for single-mode Gaussian channels
in terms of the relevant channel parameters. (i)~Thermal-loss channel
$\mathcal{E}_{\text{loss}}$ ($g<1$), studied in Ref.~\cite{QKDpaper}, and
amplifier channel $\mathcal{E}_{\text{amp}}$ ($g>1$) with $\bar{n}=1$ thermal
photon. The upper bounds in Eqs.~(\ref{LossUB}) and~(\ref{ampliBOUND}) (red
solid lines) are compared with the lower bounds (blue solid lines) obtained
with the (reverse) coherent information of these channels, i.e.,
$I_{RC}(\mathcal{E}_{\text{loss}})$~\cite{ReverseCAP} and $I_{C}%
(\mathcal{E}_{\text{amp}})~$\cite{HolevoWerner}. The two-way assisted
capacities are contained in the shadowed areas. We also show the upper bounds
derived in Ref.~\cite{TGW,Squash3} (dotted) and Ref.~\cite{GEW} (dashed) based
on the squashed entanglement. (ii)~Additive-noise channel $\mathcal{E}%
_{\text{add}}$ with classical Gaussian noise $\xi$. We compare our upper bound
of Eq.~(\ref{AdditiveUB}) with the best lower bound $I_{C}(\mathcal{E}%
_{\text{add}})$~\cite{HolevoWerner}. We also show the previous upper bounds of
Ref.~\cite{Squash3} (dotted), Ref.~\cite{GEW} (dashed), and
Ref.~\cite{HolevoWerner} (dot-dashed).~ ~~}%
\label{FigTOTAL}%
\end{figure}

Now consider the additive-noise Gaussian channel $\mathcal{E}_{\text{add}}$,
whose action is $\mathbf{\hat{x}}\rightarrow\mathbf{\hat{x}}+(z,z)^{T}$ where
$z$ is a classical Gaussian variable with zero mean and variance $\xi\geq0$.
For this channel we find (see Methods)
\begin{equation}
\Phi(\xi)\leq\frac{\xi-1}{\ln2}-\log_{2}\xi~~\text{for~}\xi<1\text{,}
\label{AdditiveUB}%
\end{equation}
and $\Phi(\xi)=0$ otherwise. The best lower bound is its coherent information
$I_{C}(\mathcal{E}_{\text{add}})=-1/\ln2-\log_{2}\xi$~\cite{HolevoWerner}. In
Fig.~\ref{FigTOTAL} we explicitly show that our upper bounds, computed from
the relative entropy of entanglement, are the tightest in the literature,
e.g., compared with previous results in
Refs.~\cite{TGW,Squash3,GEW,HolevoWerner}. The two-way assisted capacities
($K$ and $Q_{2}=D_{2}$) of these Gaussian channels are in the shadowed areas.
For both $\mathcal{E}_{\text{loss}}$ and $\mathcal{E}_{\text{amp}}$, these
areas shrink to a single line for $\bar{n}=0$. 

It is interesting to note how quantum communication rapidly degrades when we
compose quantum channels. For instance, a quantum-limited amplifier with gain
$2$ can transmit $Q_{2}=1$ qubit per use from Alice to Bob. This is the same
amount which can be transmitted from Bob to Charlie, through a pure-loss
channel with transmissivity $1/2$. By using Charlie as a quantum repeater,
Alice can therefore transmit at least $1$ qubit per use to Bob. If we remove
Charlie, and we compose the two channels, we obtain an additive-noise Gaussian
channel with variance $\xi=1/2$, for which $Q_{2}\lesssim0.278$ qubits per use.

\subsection{Discrete-variable channels}

We now study the ultimate limit for secret key generation, entanglement
distillation and quantum communication through basic channels in finite
dimension. Let us start with the qubit erasure channel which is defined as
\begin{equation}
\mathcal{E}_{\text{erase}}(\rho)=(1-p)\rho+p|2\rangle\langle2|,
\end{equation}
where $\langle2|\rho|2\rangle=0$ for any input $\rho$. In the Methods, we
compute its entanglement flux $\Phi(\mathcal{E}_{\text{erase}})\leq1-p$.
Because $Q_{2}(\mathcal{E}_{\text{erase}})=1-p$~\cite{ErasureChannel} and
$Q_{2}\leq K\leq\Phi$, this automatically establishes its secret-key capacity
\begin{equation}
K(\mathcal{E}_{\text{erase}})=1-p~.
\end{equation}

Let us compute the entanglement flux of a generic Pauli channel $\mathcal{P}$
acting on qubits. This can be written as
\begin{equation}
\mathcal{P}(\rho)=\sum_{k}p_{k}P_{k}\rho P_{k},
\end{equation}
where $P_{k}\in\{I,X,Y,Z\}$ are Pauli operators~\cite{WildeBOOK,BoundPRE} and
$\mathbf{p}=\{p_{k}\}$ is a probability distribution. It is known that its
(reverse) coherent information is $I_{(R)C}(\mathcal{P})=1-H(\mathbf{p})$,
where $H(\cdot)$ is the Shannon entropy. We prove that
\begin{equation}
\Phi(\mathcal{P})\leq1+H_{2}(p_{1}+p_{2})-H(\mathbf{p})~, \label{Pauligen}%
\end{equation}
where $H_{2}(\cdot)$ is the binary Shannon entropy (see Methods). We now
specialize this result for the depolarizing and dephasing channels.

The depolarizing channel is a qubit Pauli channel with probability
distribution $\mathbf{p}=\{1-3p/4,p/4,p/4,p/4\}$. For this channel, we have
$\Phi(\mathcal{P}_{\text{depol}})\leq f(p)$ where%
\begin{equation}
f(p):=1+H_{2}\left(  \frac{p}{2}\right)  -H_{2}\left(  \frac{3p}{4}\right)
-\frac{3p}{4}\log_{2}3~,
\end{equation}
which may be improved into
\begin{equation}
\Phi(\mathcal{P}_{\text{depol}})\leq\min_{\epsilon}(1-\alpha)f(\epsilon)~,
\end{equation}
where $\alpha=(p-\epsilon)/(2/3-\epsilon)$ and $0\leq\epsilon\leq p\leq2/3$
(see Methods). This bound is close to (but do not improve) that derived from
the squashed entanglement~\cite{GEW}.

Finally, consider the dephasing channel $\mathcal{P}_{\text{deph}}$, which is
a qubit Pauli channel with $\mathbf{p}=\{p,0,0,1-p\}$. From
Eq.~(\ref{Pauligen}) we derive $\Phi\leq1-H_{2}(p)=I_{C}$, which implies
\begin{equation}
Q_{2}(\mathcal{P}_{\text{deph}})=K(\mathcal{P}_{\text{deph}})=1-H_{2}(p)~.
\end{equation}
Thus we determine the two-way capacities of this channel, which turn out to
coincide with its unassisted quantum capacity $Q$~\cite{degradable}. This
result can be extended to arbitrary dimension $d$.

Consider a qudit with basis $\{\left\vert j\right\rangle \}$\ with
$j=0,\ldots,d-1$. The generalized phase operator is $Z\left\vert
j\right\rangle =\omega^{j}\left\vert j\right\rangle $ with $\omega:=\exp
(i2\pi/d)$, and the generalized dephasing channel with probability $p$ has
Kraus representation
\begin{equation}
\mathcal{P}_{\text{deph}}^{d}(\rho)=\sum_{m=0}^{d-1}P_{m}Z^{m}\rho
(Z^{m})^{\dagger},
\end{equation}
where $P_{m}:=\binom{d-1}{m}p^{m}(1-p)^{d-1-m}$. Let us set $\mathbf{P}%
=\{P_{m}\}$, then we find (see Methods)%
\begin{equation}
Q_{2}(\mathcal{P}_{\text{deph}}^{d})=K(\mathcal{P}_{\text{deph}}^{d})=\log
_{2}d-H(\mathbf{P}).
\end{equation}

\subsection{Conclusions}

In this work we have studied the ultimate rates of quantum communication,
entanglement distillation and key distribution between two parties who are
connected by a stretchable channel, which suitably commutes with
teleportation. Thanks to this property we have been able to reduce the most
general two-way assisted protocol based on adaptive LOCCs into a much simpler
non-adaptive protocol, where each use of the channel is mapped into its Choi
matrix, and the adaptive LOCCs are collapsed into a single final LOCC.

This simplification allowed us to exploit basic properties of the relative
entropy of entanglement, and to compute the entanglement flux of the channel,
which provides an upper bound for the various two-way assisted capacities.
Remarkably, this upper bound turned out to coincide with known lower bounds
for several important quantum channels, such the quantum-limited bosonic
amplifier, the dephasing channel in arbitrary dimension and the qubit erasure
channel, for which the various two-way assisted capacities are now fully established.

In the absence of pre-shared entanglement, the entanglement flux of a
stretchable channel can only be surpassed by using a quantum repeater. For
this reason, our results provide the most general and precise benchmarks for
testing the rate performance of quantum repeaters in both discrete- and
continous-variable settings.

\textbf{Acknowledgments}.~This work has been supported by the
EPSRC\ via the `UK Quantum Communications HUB' (EP/M013472/1) and
`qDATA' (EP/L011298/1). S.P. would like to thank S. L. Braunstein, G. Spedalieri, and M. M. Wilde for comments.

%\newpage

\bigskip

\section{Methods}

\subsection{Ideal teleportation}

Ideal teleportation is based on the use of an ideal EPR state $\Phi
_{AA^{\prime}}^{\text{EPR}}$ of systems $A$\ and $A^{\prime}$. For a qudit of
arbitrary dimension $d$, this is a generalized Bell state, i.e., maximally
entangled state of the form
\begin{equation}
\Phi_{AA^{\prime}}^{\text{EPR}}=d^{-1/2}\sum_{i=1}^{d}\left\vert
i\right\rangle _{A}\left\vert i\right\rangle _{A^{\prime}}~. \label{EPRfinite}%
\end{equation}
In particular, it is the usual Bell state $(\left\vert 00\right\rangle
+\left\vert 11\right\rangle )/\sqrt{2}$ for a qubit. For CVs, one has to take
the limit of $d\rightarrow+\infty$ in Eq.~(\ref{EPRfinite}); this is a
two-mode squeezed vacuum state~\cite{RMP} with infinite energy, i.e.,
realizing $\hat{q}_{A}=\hat{q}_{A^{\prime}}$ for position and $\hat{p}%
_{A}=-\hat{p}_{A^{\prime}}$ for momentum. Such unbounded state must always be
considered in this limit.

Correspondingly, we consider an ideal Bell detection acting on systems $a$ and
$A$, which is a projection on Bell states $\Phi_{aA}^{k}$ where the label $k$
(outcome of the measurement) takes $2^{d}$ possible values for qudits, while
it is complex for CVs~\cite{telereview}. More precisely, the ideal Bell
detection is a POVM with generic measurement operator
\begin{equation}
\Phi_{aA}^{k}:=(T_{k}^{a}\otimes I^{A})^{\dagger}\Phi_{aA}^{\text{EPR}}%
(T_{k}^{a}\otimes I^{A})~,
\end{equation}
where $T_{k}$ is a teleportation unitary. At any dimension $d$, we call
teleportation set $\mathcal{S}_{d}$, the set of all teleportation unitaries.
For $d<+\infty$ (qudit), $\mathcal{S}_{d}$ is composed by $d^{2}$ generalized
Pauli operators. For $d=+\infty$ (CV system), $\mathcal{S}_{\infty}$ is
composed by infinite displacement operators~\cite{RMP}.

Let us better characterize the teleportation set. Denote by $\{|j\rangle\}$
the computational basis of a qudit, with $j\in\mathbb{Z}_{d}:=\{0,\ldots
,d-1\}$. Any qudit unitary can be expanded in terms of $d^{2}$ generalized
Pauli operators $X^{a}Z^{b}$ with $a,b\in\mathbb{Z}_{d}$. These are defined by
the following unitary (non-Hermitian) operators%
\begin{equation}
X\left\vert j\right\rangle =\left\vert j\oplus1\right\rangle ~,~Z\left\vert
j\right\rangle =\omega^{j}\left\vert j\right\rangle ~, \label{Pauli_DEF}%
\end{equation}
where $\oplus$ is the modulo $d$ addition and $\omega:=\exp(i2\pi/d)$. Thus we
have $\mathcal{S}_{d}=\{X^{a}Z^{b}\}$ with $a,b\in\mathbb{Z}_{d}$. Note that,
from $\mathcal{S}$, we may construct the set of finite-dimensional
displacement operators $D(j,a,b):=\omega^{j}X^{a}Z^{b}$ with $j,a,b\in
\mathbb{Z}_{d}$ which forms the finite-dimensional Weyl-Heisenberg group (or
Pauli group). For instance, for a qubit ($d=2$), we have $\mathcal{S}%
_{2}=\{I,X,XZ,Z\}$ and the group $\pm1\times\left\{  I,X,XZ,Z\right\}  $. For
CV systems, $\mathcal{S}_{\infty}=\{D(k)\}$ with $k$ complex and $D(k)$ being
a displacement operator~\cite{RMP}. This is the infinite-dimensional
Weyl-Heisenberg group.

Given an arbitrary state $\rho$ on some input system $a$, this is perfectly
teleported onto $A^{\prime}$ by the following procedure. System $a$ and EPR
system $A$ are subject to Bell detection. For any given outcome $k$, the other
EPR system $A^{\prime}$ is projected onto $T_{k}\rho T_{k}^{\dagger}$ where
$T_{k}\in\mathcal{S}_{d}$. The last step is the CC of the outcome $k$, which
allows one to undo the teleportation unitary by applying $T_{k}^{\dagger}$ to
system $A^{\prime}$. Note that this process also teleports all correlations
that the input system $a$ may have with other ancillary systems. In the
following Methods' section we consider the presence of a channel and the full
mathematics of the stretching mechanism.

\subsection{Detailed maths of teleportation stretching}

Refer to the scenario depicted in panel~(ii) of Fig.~\ref{panpic}, where we
insert an ideal teleportation circuit before the use of the channel
$\mathcal{E}$. For simplicity drop the index from the transmitted systems, so
that we have an ideal EPR\ state $\Phi_{AA^{\prime}}^{\text{EPR}}$\ of systems
$A$ and $A^{\prime}$ and an ideal Bell detection of systems $a$ and $A$.
System $A^{\prime}$ is transformed into the output $B$ by the action of the
channel. Let us consider the initial state $\rho_{\mathbf{a}a\mathbf{b}}$
coming from the application of the first LOCC $\rho_{\mathbf{a}a\mathbf{b}%
}=\Lambda_{1}(\sigma_{\mathbf{a}}\otimes\sigma_{\mathbf{b}})$ on Alice's and
Bob's local states $\sigma_{\mathbf{a}}$ and $\sigma_{\mathbf{b}}$. Including
the ideal EPR state, we have the global state $\rho_{\mathbf{a}a\mathbf{b}%
}\otimes\Phi_{AA^{\prime}}^{\text{EPR}}$. Performing the Bell detection on
systems $a$ and $A$, with outcome $\Phi_{aA}^{k}$ and probability $p_{k}$,
leads to the\ global state
\begin{gather}
\rho_{\mathbf{a}a\mathbf{b}AA^{\prime}}^{k}:=\Phi_{aA}^{k}(\rho_{\mathbf{a}%
a\mathbf{b}}\otimes\Phi_{AA^{\prime}}^{\text{EPR}})\Phi_{aA}^{k\dagger
}\label{provaP}\\
=\Phi_{aA}^{k}\otimes\rho_{\mathbf{a}A\mathbf{^{\prime}b}}^{k}~.
\label{provaPP}%
\end{gather}
where
\begin{equation}
\rho_{\mathbf{a}A\mathbf{^{\prime}b}}^{k}=T_{k}^{A^{\prime}}\rho
_{\mathbf{a}A^{\prime}\mathbf{b}}T_{k}^{A^{\prime}\dagger}~. \label{porvaS}%
\end{equation}
For simplicity of notation, we omit identities when they are involved in
tensor products with other operators.

\textbf{Proof.}~From the definition of the previous global state and using
$\Phi_{aA}^{k}=\left\vert \Phi^{k}\right\rangle _{aA}\left\langle \Phi
^{k}\right\vert $, we get%
\begin{align*}
\rho_{\mathbf{a}a\mathbf{b}AA^{\prime}}^{k}  &  =\left\vert \Phi
^{k}\right\rangle _{aA}\left\langle \Phi^{k}\right\vert (\rho_{\mathbf{a}%
a\mathbf{b}}\otimes\Phi_{AA^{\prime}}^{\text{EPR}})\left\vert \Phi
^{k}\right\rangle _{aA}\left\langle \Phi^{k}\right\vert \\
&  =\Phi_{aA}^{k}\otimes\rho_{\mathbf{a}A\mathbf{^{\prime}b}}^{k}~,
\end{align*}
where
\begin{align}
\rho_{\mathbf{a}A\mathbf{^{\prime}b}}^{k}  &  :=_{aA}\left\langle
\Phi^{\text{EPR}}\right\vert T_{k}^{a}~\rho_{\mathbf{a}a\mathbf{b}}\nonumber\\
&  \otimes\left\vert \Phi^{\text{EPR}}\right\rangle _{AA^{\prime}}\left\langle
\Phi^{\text{EPR}}\right\vert T_{k}^{a\dagger}\left\vert \Phi^{\text{EPR}%
}\right\rangle _{aA}~.\nonumber
\end{align}
Up to normalizations, we may write
\begin{align*}
\rho_{\mathbf{a}A\mathbf{^{\prime}b}}^{k}  &  =\sum_{ijlm}~_{aA}\left\langle
ii\right\vert T_{k}^{a}~\rho_{\mathbf{a}a\mathbf{b}}\left\vert jj\right\rangle
_{AA^{\prime}~AA^{\prime}}\left\langle ll\right\vert T_{k}^{a\dagger
}\left\vert mm\right\rangle _{aA}\\
&  =\sum_{il}~_{a}\left\langle i\right\vert T_{k}^{a}~\rho_{\mathbf{a}%
a\mathbf{b}}~T_{k}^{a\dagger}\left\vert l\right\rangle _{a}\otimes\left\vert
i\right\rangle _{A^{\prime}}\left\langle l\right\vert \\
&  =M_{a\rightarrow A^{\prime}}~T_{k}^{a}~\rho_{\mathbf{a}a\mathbf{b}}%
~T_{k}^{a\dagger}~M_{a\rightarrow A^{\prime}}^{\dagger}\\
&  =T_{k}^{A^{\prime}}~\rho_{\mathbf{a}A^{\prime}\mathbf{b}}~T_{k}^{A^{\prime
}\dagger},
\end{align*}
where $M_{a\rightarrow A^{\prime}}:=\sum_{i}\left\vert i\right\rangle
_{A^{\prime}~a}\left\langle i\right\vert $ maps $a$ into system $A^{\prime}%
$.~$\blacksquare$

Now let us apply channel $\mathcal{E}_{A^{\prime}}$ to the conditional global
state $\rho_{\mathbf{a}a\mathbf{b}AA^{\prime}}^{k}=\Phi_{aA}^{k}\otimes
\rho_{\mathbf{a}A\mathbf{^{\prime}b}}^{k}$. Using Eq.~(\ref{porvaS}) we
derive
\begin{align}
\rho_{\mathbf{a}B\mathbf{b}}^{k}  &  =\mathcal{E}_{A^{\prime}}\left[
T_{k}^{A^{\prime}}\rho_{\mathbf{a}A^{\prime}\mathbf{b}}T_{k}^{A^{\prime
}\dagger}\right] \nonumber\\
&  =U_{k}^{B}~\mathcal{E}_{A^{\prime}}(\rho_{\mathbf{a}A^{\prime}\mathbf{b}%
})~U_{k}^{B\dagger}~,
\end{align}
where we have used the fact that $\mathcal{E}$ is stretchable. Then, Bob
applies the inverse unitary $U_{k}^{B\dagger}$ which provides
\begin{equation}
\rho_{\mathbf{a}B\mathbf{b}}=\mathcal{E}_{A^{\prime}}(\rho_{\mathbf{a}%
A^{\prime}\mathbf{b}})=\mathcal{E}_{a}(\rho_{\mathbf{a}a\mathbf{b}})~,
\label{reduceS}%
\end{equation}
where we have re-labeled $A^{\prime}\rightarrow a$ in the last equality.

Note that the output state $\rho_{\mathbf{a}B\mathbf{b}}$ is independent on
the outcome $k$ of the Bell detection and corresponds to the output state that
one would achieve by direct transmission of system $a$ through the channel, as
depicted in panel (i) of Fig.~\ref{panpic}. The final step is the LOCC
$\Lambda_{2}$ which provides
\[
\rho_{\mathbf{ab}}=\Lambda_{2}(\rho_{\mathbf{a}B\mathbf{b}})~.
\]
As a matter of fact, the second LOCC is globally described by $\Lambda_{2}%
^{k}=\Lambda_{2}\circ\mathcal{U}_{k}^{B}$ where $\mathcal{U}_{k}^{B}%
(\rho)=U_{k}^{B\dagger}\rho U_{k}^{B}$.

Now let us apply the same operations to $\rho_{\mathbf{a}a\mathbf{b}%
AA^{\prime}}^{k}$ when it is written in the equivalent form of
Eq.~(\ref{provaP}). After the channel we have
\begin{align*}
\mathcal{E}_{A^{\prime}}\left(  \rho_{\mathbf{a}a\mathbf{b}AA^{\prime}}%
^{k}\right)   &  =\mathcal{E}_{A^{\prime}}\left[  \Phi_{aA}^{k}(\rho
_{\mathbf{a}a\mathbf{b}}\otimes\Phi_{AA^{\prime}}^{\text{EPR}})\Phi
_{aA}^{k\dagger}\right] \\
&  =\Phi_{aA}^{k}\left[  \rho_{\mathbf{a}a\mathbf{b}}\otimes\mathcal{E}%
_{A^{\prime}}\left(  \Phi_{AA^{\prime}}^{\text{EPR}}\right)  \right]
\Phi_{aA}^{k\dagger}\\
&  =\Phi_{aA}^{k}\left(  \rho_{\mathbf{a}a\mathbf{b}}\otimes\rho_{\mathcal{E}%
}^{AB}\right)  \Phi_{aA}^{k\dagger}\\
&  =\mathcal{B}_{aA}^{k}\left(  \rho_{\mathbf{a}a\mathbf{b}}\otimes
\rho_{\mathcal{E}}^{AB}\right)  ,
\end{align*}
where $\rho_{\mathcal{E}}^{AB}$ is the Choi matrix of the channel
$\mathcal{E}$ and $\mathcal{B}_{aA}^{k}(\rho):=\Phi_{aA}^{k}\rho\Phi
_{aA}^{k\dagger}$. Then, Bob applies the conditional LO $\mathcal{U}_{k}^{B}$
so that we get
\[
\rho_{\mathbf{a}a\mathbf{b}AB}^{k}=\mathcal{U}_{k}^{B}\circ\mathcal{B}%
_{aA}^{k}\left(  \rho_{\mathbf{a}a\mathbf{b}}\otimes\rho_{\mathcal{E}}%
^{AB}\right)  ,
\]

By tracing over systems $a$ and $A$, we must retrieve the reduced state in
Eq.~(\ref{reduceS}), which does not depend on $k$, i.e., we have%
\[
\rho_{\mathbf{a}B\mathbf{b}}=\mathrm{Tr}_{aA}\left[  \mathcal{U}_{k}^{B}%
\circ\mathcal{B}_{aA}^{k}\left(  \rho_{\mathbf{a}a\mathbf{b}}\otimes
\rho_{\mathcal{E}}^{AB}\right)  \right]  .
\]
Finally, we apply the LOCC $\Lambda_{2}$ to systems $\mathbf{a}B\mathbf{b}$,
so that we have%
\begin{align}
\rho_{\mathbf{ab}}  &  =\Lambda_{2}(\rho_{\mathbf{a}B\mathbf{b}}%
)=\mathrm{Tr}_{aA}\left[  \Lambda_{2}^{k}\circ\mathcal{B}_{aA}^{k}\left(
\rho_{\mathbf{a}a\mathbf{b}}\otimes\rho_{\mathcal{E}}^{AB}\right)  \right]
\nonumber\\
&  =\mathrm{Tr}_{aA}\left\{  \Lambda_{2}^{k}\circ\mathcal{B}_{aA}^{k}\left[
\Lambda_{1}(\sigma_{\mathbf{a}}\otimes\sigma_{\mathbf{b}})\otimes
\rho_{\mathcal{E}}^{AB}\right]  \right\}  . \label{pannello3}%
\end{align}
The last expression in Eq.~(\ref{pannello3}) describes the stretched scenario
in the panel~(iii) of Fig.~\ref{panpic}.

It is clear that $\rho_{\mathbf{ab}}$ does not depend on $k$. This means that
it is equal to the mean state obtained by averaging over the Bell outcomes. In
other words, we may write%
\begin{equation}
\rho_{\mathbf{ab}}=\sum_{k}p_{k}\rho_{\mathbf{ab}}=\mathrm{Tr}_{aA}\left[
\Delta\left(  \rho_{\mathbf{a}a\mathbf{b}}\otimes\rho_{\mathcal{E}}%
^{AB}\right)  \right]  , \label{abquasi}%
\end{equation}
where%
\begin{equation}
\Delta:=\sum_{k}p_{k}(\Lambda_{2}^{k}\circ\mathcal{B}_{aA}^{k})
\end{equation}
is a trace-preserving LOCC. \ In the first transmission, the state
$\rho_{\mathbf{a}a\mathbf{b}}$ is prepared locally, so that we may simplify
Eq.~(\ref{abquasi}) into the following%
\begin{equation}
\rho_{\mathbf{ab}}=\bar{\Lambda}(\rho_{\mathcal{E}}^{AB}),
\end{equation}
where $\bar{\Lambda}$ is a trace-preserving LOCC. This is the scenario
depicted in panel~(iv) of Fig.~\ref{panpic}.

\subsubsection{Iteration rule}

We can easily show that, by iteration, we can stretch all the
transmission instances of the adaptive protocol, as in panel~(v)
of Fig.~\ref{panpic}. Let us introduce the label $i=1,2,\ldots,n$
to denote the various systems and operations associated with the
$i$th
transmission. We can modify Eq.~(\ref{abquasi}) into%
\begin{equation}
\rho_{\mathbf{ab}}^{i}=\mathrm{Tr}_{a_{i}A_{i}}\left[  \Delta_{i}\left(
\rho_{\mathbf{a}a_{i}\mathbf{b}}\otimes\rho_{\mathcal{E}}^{A_{i}B_{i}}\right)
\right]  \label{ite1}%
\end{equation}
where
\begin{equation}
\rho_{\mathbf{a}a_{i}\mathbf{b}}=\rho_{\mathbf{ab}}^{i-1},~~~\rho
_{\mathbf{ab}}^{0}=\Lambda_{1}(\sigma_{\mathbf{a}}\otimes\sigma_{\mathbf{b}}).
\label{ite2}%
\end{equation}

Thus, for $n=2$ transmissions, we may write%
\begin{align}
\rho_{\mathbf{ab}}^{2}  &  =\mathrm{Tr}_{a_{2}A_{2}}\left[  \Delta_{2}\left(
\rho_{\mathbf{a}a_{2}\mathbf{b}}\otimes\rho_{\mathcal{E}}^{A_{2}B_{2}}\right)
\right] \nonumber\\
&  =\mathrm{Tr}_{a_{2}A_{2}}\left[  \Delta_{2}\left(  \rho_{\mathbf{ab}}%
^{1}\otimes\rho_{\mathcal{E}}^{A_{2}B_{2}}\right)  \right] \nonumber\\
&  =\mathrm{Tr}_{a_{2}A_{2}}\left\{  \Delta_{2}\left\{  \mathrm{Tr}%
_{a_{1}A_{1}}\left[  \Delta_{1}\left(  \rho_{\mathbf{ab}}^{0}\otimes
\rho_{\mathcal{E}}^{A_{1}B_{1}}\right)  \right]  \otimes\rho_{\mathcal{E}%
}^{A_{2}B_{2}}\right\}  \right\} \nonumber\\
&  \overset{(\ast)}{=}\mathrm{Tr}_{a_{1}a_{2}A_{1}A_{2}}\left\{  \Delta
_{2}\left[  \Delta_{1}\left(  \rho_{\mathbf{ab}}^{0}\otimes\rho_{\mathcal{E}%
}^{A_{1}B_{1}}\right)  \otimes\rho_{\mathcal{E}}^{A_{2}B_{2}}\right]  \right\}
\nonumber\\
&  \overset{(\ast)}{=}\mathrm{Tr}_{a_{1}a_{2}A_{1}A_{2}}\left[  \Delta
_{2}\circ\Delta_{1}\left(  \rho_{\mathbf{ab}}^{0}\otimes\rho_{\mathcal{E}%
}^{A_{1}B_{1}}\otimes\rho_{\mathcal{E}}^{A_{2}B_{2}}\right)  \right]
\label{lastED}%
\end{align}
where in $(\ast)$ we exploit the fact that the LOCC $\Delta_{i}$ acts on
systems $\mathbf{ab}a_{i}A_{i}B_{i}$. Because $\rho_{\mathbf{ab}}^{0}$ is
prepared locally, we may simplified Eq.~(\ref{lastED}) into%
\begin{equation}
\rho_{\mathbf{ab}}^{2}=\bar{\Lambda}\left(  \rho_{\mathcal{E}}^{A_{1}B_{1}%
}\otimes\rho_{\mathcal{E}}^{A_{2}B_{2}}\right)  .
\end{equation}
By iterating $n$ times Eqs.~(\ref{ite1}) and~(\ref{ite2}), we derive%
\begin{align}
\rho_{\mathbf{ab}}^{n}  &  =\mathrm{Tr}_{a_{1}\ldots a_{n}A_{1}\ldots A_{n}%
}\left[  \Delta_{n}\circ\cdots\circ\Delta_{1}\left(  \rho_{\mathbf{ab}}%
^{0}\otimes%
%TCIMACRO{\tbigotimes \nolimits_{i=1}^{n}}%
%BeginExpansion
{\textstyle\bigotimes\nolimits_{i=1}^{n}}
%EndExpansion
\rho_{\mathcal{E}}^{A_{i}B_{i}}\right)  \right] \nonumber\\
&  =\bar{\Lambda}\left(  \rho_{\mathcal{E}}^{\otimes n}\right)  ~.
\end{align}

\subsection{(Reverse) coherent information of a channel}

Consider a quantum channel $\mathcal{E}$ which is applied to some input state
$\rho_{A}$ of system $A$. Let us introduce an auxiliary system $R$ and
consider the purification $|\psi\rangle_{RA}$ of $\rho_{A}$. We can therefore
consider the extended channel $\rho_{RB}=(\mathcal{I}\otimes\mathcal{E}%
)(|\psi\rangle\langle\psi|)$. By definition, the coherent information for
channel $\mathcal{E}$ and the input state $\rho_{A}$ is~\cite{Schu96,Lloyd97}%
\begin{equation}
I_{C}(\mathcal{E},\rho_{A})=S(\rho_{B})-S(\rho_{RB})~,
\end{equation}
where $S(\cdot)$ is the von Neumann entropy and $\rho_{B}=\mathrm{Tr}_{R}%
(\rho_{RB})$. This is also denoted as
\begin{equation}
I_{C}(\mathcal{E},\rho_{A})=I(A\rangle B)_{\rho_{RB}}~.
\end{equation}
Similarly, the reverse coherent information is~\cite{RevCohINFO,ReverseCAP}%
\begin{equation}
I_{RC}(\mathcal{E},\rho_{A})=S(\rho_{R})-S(\rho_{RB})~,
\end{equation}
where $\rho_{R}=\mathrm{Tr}_{B}(\rho_{RB})$. This is also denoted as
\begin{equation}
I_{RC}(\mathcal{E},\rho_{A})=I(A\langle B)_{\rho_{RB}}~.
\end{equation}

When the input state $\rho_{A}$ is a maximally-mixed state, its purification
is a maximally-entangled state $\Phi_{RA}^{\text{EPR}}$, so that $\rho_{RB}$
is the Choi-matrix of the channel, i.e., $\rho_{\mathcal{E}}$. We then define
the coherent information of the channel as
\begin{equation}
I_{C}(\mathcal{E})=I(A\rangle B)_{\rho_{\mathcal{E}}}~.
\end{equation}
Similarly, its reverse coherent information is
\begin{equation}
I_{RC}(\mathcal{E})=I(A\langle B)_{\rho_{\mathcal{E}}}~.
\end{equation}

These quantities are computable in any dimension and they are achievable rates
for one-way assisted entanglement distillation, according to the hashing
inequality~\cite{DWrates} which also applies to energy-constrained states in
infinite dimension and suitable limits for infinite energy~\cite{QKDpaper}. It
is clear that $I_{C}(\mathcal{E})$ is a lower bound for the one-shot
(unassisted or forward-assisted) quantum capacity of the channel, i.e.,%
\begin{equation}
I_{C}(\mathcal{E})\leq Q^{(1)}(\mathcal{E})=\max_{|\psi\rangle}I_{C}%
(\mathcal{E},\rho_{A})~.
\end{equation}
Indeed, one may have $I_{C}(\mathcal{E})=Q^{(1)}(\mathcal{E})$, for instance
for the pure-loss channel~\cite{HolevoWerner,Wolf}.

Note that for unital channels, i.e., channels preserving the identity
$\mathcal{E}(I)=I$, we have $I_{C}(\mathcal{E})=I_{RC}(\mathcal{E})$. This is
just a consequence of the fact that, the reduced states $\rho_{A}$ and
$\rho_{R}$ of a maximally entangled state $\Phi_{RA}^{\text{EPR}}$ is a
maximally-mixed state $I/d$, where $d$ is the dimension of the Hilbert space
(including the limit for $d\rightarrow+\infty$). If the channel is unital,
also the reduced output state $\rho_{B}=\mathcal{E}(\rho_{A})$ is
maximally-mixed. As a result, $S(\rho_{B})=S(\rho_{A})=S(\rho_{R})$ and we may
write $I_{C}(\mathcal{E})=I_{RC}(\mathcal{E}):=I_{(R)C}(\mathcal{E})$. In the
specific case of discrete variable systems ($d<+\infty$), we have $S(\rho
_{R})=\log_{2}d$ and we may write%
\begin{equation}
I_{(R)C}(\mathcal{E})=\log_{2}d-S(\rho_{\mathcal{E}})~. \label{unitalGEN}%
\end{equation}
In particular, for qubits ($d=2$), one has%
\begin{equation}
I_{(R)C}(\mathcal{E})=1-S(\rho_{\mathcal{E}})~. \label{unitalRC}%
\end{equation}

\subsection{Convexity of the entanglement flux}

This property is inherited from the convexity of the relative entropy of
entanglement. For any ensemble of states $\{p_{i},\rho_{i}\}$ with average
$\bar{\rho}=\sum_{i}p_{i}\rho_{i}$, one has~\cite{Donald}%
\begin{equation}
E_{R}(\bar{\rho})\leq\sum_{i}p_{i}E_{R}(\rho_{i})~.
\end{equation}
Let us consider the mean channel $\overline{\mathcal{E}}=\sum_{i}%
p_{i}\mathcal{E}_{i}$, defined for an ensemble of channels $\{p_{i}%
,\mathcal{E}_{i}\}$. In terms of Choi matrices, we have%
\begin{align*}
\rho_{\overline{\mathcal{E}}}  &  :=(I\otimes\overline{\mathcal{E}})\left(
\Phi^{\text{EPR}}\right) \\
&  =\sum_{i}p_{i}(I\otimes\mathcal{E}_{i})\left(  \Phi^{\text{EPR}}\right)
=\sum_{i}p_{i}\rho_{\mathcal{E}_{i}}~.
\end{align*}
As a result, we may write%
\begin{equation}
\Phi(\overline{\mathcal{E}}):=E_{R}(\rho_{\overline{\mathcal{E}}})\leq\sum
_{i}p_{i}E_{R}(\rho_{\mathcal{E}_{i}})=\sum_{i}p_{i}\Phi(\mathcal{E}_{i}).
\end{equation}

\subsection{Entanglement flux of a canonical form}

Let us consider a single-mode Gaussian channel. By means of local unitaries
this channel can always be put in canonical form~\cite{RMP} whose general
action on input quadratures $\mathbf{\hat{x}}=(\hat{q},\hat{p})^{T}$ is given
by%
\begin{equation}
\mathbf{\hat{x}}\rightarrow\mathbf{G\hat{x}}+\mathbf{H\hat{x}}_{E(\bar{n}%
)}+(z,z)^{T}~, \label{genFORM}%
\end{equation}
where $\mathbf{G}$ and $\mathbf{H}$ are diagonal matrices, $E(\bar{n})$ is an
environmental thermal mode with $\bar{n}$ mean photons, and $z$ is a classical
Gaussian variable with zero mean and variance $\xi\geq0$. Depending on the
specific form (thermal-loss channel, amplifier etc...) we have different
expressions in Eq.~(\ref{genFORM}). For instance, the thermal loss channel has
$\mathbf{G}=\sqrt{g}\mathbf{I}$, $\mathbf{H}=\sqrt{1-g}\mathbf{I}$\ with
$0\leq g\leq1$ and $\xi=0$, so that its action is\ $\mathbf{\hat{x}%
}\rightarrow\sqrt{g}\mathbf{\hat{x}}+\sqrt{1-g}\mathbf{\hat{x}}_{E(\bar{n})}$.
All these channels are clearly stretchable. In fact, the effect of an ideal CV
teleportation is the $k$-dependent phase-space displacement of the
input\ $\mathbf{\hat{x}}\rightarrow\mathbf{\hat{x}}+\mathbf{d}_{k}$, which is
just mapped into $\mathbf{d}_{k}\rightarrow\mathbf{Gd}_{k}$.

Since bosonic systems have an $\infty$-dimensional Hilbert space, for a
canonical form $\mathcal{E}$ we need to compute%
\begin{equation}
\Phi(\mathcal{E})=E_{R}\left(  \rho_{\mathcal{E}}\right)  \leq\lim\inf_{\mu
}E_{R}\left(  \rho_{\mu}\right)  ~,
\end{equation}
where $\rho_{\mu}:=(\mathcal{I}\otimes\mathcal{E})(\Phi_{\mu}^{\text{EPR}})$
and $\Phi_{\mu}^{\text{EPR}}$ is a two-mode squeezed vacuum state with
variance $\mu\geq1/2$. This is Gaussian~\cite{RMP} with covariance matrix (CM)%
\begin{equation}
V_{\mu}=\left(
\begin{array}
[c]{cc}%
\mu & c\\
c & \mu
\end{array}
\right)  \oplus\left(
\begin{array}
[c]{cc}%
\mu & -c\\
-c & \mu
\end{array}
\right)  ~, \label{EPR_CM}%
\end{equation}
where $c=\sqrt{\mu^{2}-1/4}$. One can easily check that the sequence of states
$\rho_{\mu}$ converges (in trace norm) to the target state $\rho_{\mathcal{E}%
}=(\mathcal{I}\otimes\mathcal{E})(\Phi^{\text{EPR}})$. This can be done by
compute the bound $\left\Vert \rho^{\mu}-\rho_{\mathcal{E}}\right\Vert
\leq\sqrt{1-F^{2}}$\ where $F=F(\rho_{\mu},\rho_{\mathcal{E}})$\ is the
fidelity between two Gaussian states~\cite{Banchi}. In order to bound the
entanglement flux we need to consider a suitable separable state
$\tilde{\sigma}$, so that
\begin{equation}
E_{R}\left(  \rho_{\mu}\right)  \leq S(\rho_{\mu}||\tilde{\sigma}_{\mu
})=-S(\rho_{\mu})-\mathrm{Tr}\left(  \rho_{\mu}\log_{2}\tilde{\sigma}_{\mu
}\right)  ~. \label{REcomp}%
\end{equation}
In the following, we explicitly compute the output state $\rho_{\mu}$ and a
corresponding separable state $\tilde{\sigma}_{\mu}$ for the various canonical
forms (apart from the thermal-loss channel already studied in
Ref.~\cite{QKDpaper}). These are all zero-mean Gaussian states, so that the
calculations reduce to the manipulation of their covariance matrices.

\subsubsection{Amplifier}

The amplifier channel $\mathcal{E}_{\text{amp}}$ corresponds to set
$\mathbf{G}=\sqrt{g}\mathbf{I}$, $\mathbf{H}=\sqrt{g-1}\mathbf{Z}$\ with $g>1$
and $\xi=0$ in Eq.~(\ref{genFORM}). Its action is therefore $\mathbf{\hat{x}%
}\rightarrow\sqrt{g}\mathbf{\hat{x}}+\sqrt{g-1}\mathbf{\hat{x}}_{E(\bar{n})}$.
Let us derive the output state $\rho_{\mu}:=(\mathcal{I}\otimes\mathcal{E}%
_{\text{amp}})(\Phi_{\mu}^{\text{EPR}})$. It is easy to see that this Gaussian
state has CM%
\begin{equation}
V_{\mu}^{\text{amp}}=\left(
\begin{array}
[c]{cc}%
\mu & c\sqrt{g}\\
c\sqrt{g} & \beta
\end{array}
\right)  \oplus\left(
\begin{array}
[c]{cc}%
\mu & -c\sqrt{g}\\
-c\sqrt{g} & \beta
\end{array}
\right)  ~, \label{VabOUT}%
\end{equation}
where $\beta:=g\mu+(g-1)\omega$ and $\omega:=\bar{n}+1/2$. In the limit of
$\mu\rightarrow+\infty$, this state describes the Choi matrix of the amplifier
$\rho_{\mathcal{E}_{\text{amp}}}$. Computing the minimum partially-transposed
symplectic eigenvalue of $V_{\mu}^{\text{amp}}$ and taking the limit of large
$\mu$, we can see that the Choi matrix is separable for $\bar{n}\geq
(g-1)^{-1}$ which therefore represents the entanglement-breaking threshold for
the amplifier channel. In this regime, we clearly have $\Phi(\mathcal{E}%
_{\text{amp}})=E_{R}(\rho_{\mathcal{E}_{\text{amp}}})=0$.

For $\bar{n}<(g-1)^{-1}$, we construct the separable Gaussian state
$\tilde{\sigma}_{\mu}$ with CM as in Eq.~(\ref{VabOUT}) but with the
replacement $c\sqrt{g}\rightarrow\sqrt{(\mu-1/2)(\beta-1/2)}$. The relative
entropy $S(\rho_{\mu}||\tilde{\sigma}_{\mu})$ can be computed with the formula
for Gaussian states of Ref.~\cite{QKDpaper}. In particular, up to $O(\mu
^{-1})$,\ we find the expansions%
\begin{align}
S(\rho_{\mu})  &  \rightarrow h(\bar{n})+\log_{2}e(g-1)\mu,\\
-\mathrm{Tr}\left(  \rho_{\mu}\log_{2}\tilde{\sigma}_{\mu}\right)   &
\rightarrow\frac{\ln(g\mu^{2})+2+4\omega\coth^{-1}\left(  \frac{g+1}%
{g-1}\right)  }{2\ln2}.
\end{align}
Using these in Eq.~(\ref{REcomp}) we derive the bound in Eq.~(\ref{ampliBOUND}).

\subsubsection{Conjugate of the amplifier}

Let us introduce the reflection matrix $\mathbf{Z}=\mathrm{diag}(1,-1)$. The
conjugate of the amplifier channel $\widetilde{\mathcal{E}}_{\text{amp}}$
corresponds to set $\mathbf{G}=\sqrt{-g}\mathbf{Z}$, $\mathbf{H}=\sqrt
{1-g}\mathbf{I}$\ with $g<0$ and $\xi=0$ in Eq.~(\ref{genFORM}). Its action is
therefore $\mathbf{\hat{x}}\rightarrow\sqrt{-g}\mathbf{Z\hat{x}}+\sqrt
{1-g}\mathbf{\hat{x}}_{E(\bar{n})}$. It is easy to check that the Choi matrix
$\rho_{\widetilde{\mathcal{E}}_{\text{amp}}}$ is always separable, i.e., this
channel is always entanglement-breaking, so that $\Phi(\widetilde{\mathcal{E}%
}_{\text{amp}})=0$.

\subsubsection{Additive-noise Gaussian channel}

This channel $\mathcal{E}_{\text{add}}$ corresponds to set $\mathbf{G=I}$,
$\mathbf{H=0}$ and $\xi\geq0$ in Eq.~(\ref{genFORM}). Its action is therefore
$\mathbf{\hat{x}}\rightarrow\mathbf{\hat{x}}+(z,z)^{T}$. For this channel, the
output state $\rho_{\mu}$ is Gaussian with zero-mean and CM%
\begin{equation}
V_{\mu}^{\text{add}}=\left(
\begin{array}
[c]{cc}%
\mu & c\\
c & \mu+\xi
\end{array}
\right)  \oplus\left(
\begin{array}
[c]{cc}%
\mu & -c\\
-c & \mu+\xi
\end{array}
\right)  ~. \label{Vadd}%
\end{equation}
In the limit of $\mu\rightarrow+\infty$, this state becomes the Choi matrix
$\rho_{\mathcal{E}_{\text{add}}}$ which is separable for $\xi\geq1$
(entanglement-breaking threshold for this channel). Thus, we have
$\Phi(\mathcal{E}_{\text{add}})=0$ for $\xi\geq1$. For $\xi<1$, we construct
the separable Gaussian state $\tilde{\sigma}_{\mu}$ with CM as in
Eq.~(\ref{Vadd}) but with the replacement $c\rightarrow\sqrt{(\mu-1/2)(\mu
+\xi-1/2)}$. The relative entropy $S(\rho_{\mu}||\tilde{\sigma}_{\mu})$ can be
computed with the formula for Gaussian states of Ref.~\cite{QKDpaper}. Up to
$O(\mu^{-1/2}) $,\ we find the expansions%
\begin{align}
S(\rho_{\mu})  &  \rightarrow\log_{2}(e^{2}\xi\mu),\\
-\mathrm{Tr}\left(  \rho_{\mu}\log_{2}\tilde{\sigma}_{\mu}\right)   &
\rightarrow\frac{\ln\left[  \frac{(2\mu-1)(2\xi+2\mu-1)}{4}\right]  +2(1+\xi
)}{2\ln2}.
\end{align}
By replacing in Eq.~(\ref{REcomp}) we derive the bound of
Eq.~(\ref{AdditiveUB}).

\subsubsection{Pathological forms}

There are some remaining pathological forms to consider. The $A_{2}%
$-form~\cite{RMP} is a `half' depolarizing channel and corresponds to set
$\mathbf{G}=\mathrm{diag}(1,0)$, $\mathbf{H}=\mathbf{I}$, and $\xi=0$ in
Eq.~(\ref{genFORM}). Its action is $\mathbf{\hat{x}}\rightarrow(\hat{q}%
,0)^{T}+\mathbf{\hat{x}}_{E(\bar{n})}$. It is easy to check that this is
always an entanglement-breaking channel, so that $\Phi=0$. Finally, the
$B_{1}$-form~\cite{RMP} corresponds to setting $\mathbf{G}=\mathbf{I}$,
$\mathbf{H}=\mathrm{diag}(0,1)$, $\bar{n}=0$\ and $\xi=0$ in
Eq.~(\ref{genFORM}). Its action is $\mathbf{\hat{x}}\rightarrow\mathbf{\hat
{x}}+(0,\hat{p}_{v})^{T}$ where $v$ is the vacuum. For this form we find
$\Phi=+\infty$.

\subsection{Entanglement flux of a Pauli channel}

Consider a Pauli channel $\mathcal{P}$, whose action on a quantum state $\rho$
can be written as follows
\begin{equation}
\mathcal{P}(\rho)=p_{0}\rho+p_{1}X\rho X+p_{2}Y\rho Y+p_{3}Z\rho Z,
\end{equation}
where, $p_{i}\geq0\;\forall i$, $\sum_{i}p_{i}=1$, and
\begin{equation}
X:=\left(
\begin{array}
[c]{cc}%
0 & 1\\
1 & 0
\end{array}
\right)  ,~Y:=\left(
\begin{array}
[c]{cc}%
0 & -i\\
i & 0
\end{array}
\right)  ,~Z:=\left(
\begin{array}
[c]{cc}%
1 & 0\\
0 & -1
\end{array}
\right)  .
\end{equation}
First of all let us write its Choi matrix in the computational basis. This
means that we compute%
\begin{equation}
\rho_{\mathcal{P}}=(I\otimes\mathcal{P})(\Phi^{\text{EPR}}),~\Phi^{\text{EPR}%
}=\frac{|00\rangle+|11\rangle}{\sqrt{2}}.
\end{equation}
After simple algebra we derive
\begin{equation}
\rho_{\mathcal{P}}=\frac{1}{2}\left(
\begin{array}
[c]{cccc}%
p_{0}+p3 & 0 & 0 & p_{0}-p_{3}\\
0 & p_{1}+p_{2} & p_{1}-p_{2} & 0\\
0 & p_{1}-p_{2} & p_{1}+p_{2} & 0\\
p_{0}-p3 & 0 & 0 & p_{0}+p_{3}%
\end{array}
\right)  ~. \label{Pcomp}%
\end{equation}
This state has spectral decomposition
\begin{equation}
\rho_{\mathcal{P}}=\sum_{k=0}^{3}p_{k}|k\rangle\langle k|,
\end{equation}
where the eigenvalues are the probabilities $p_{k}$ of the Pauli operators
$P_{k}\in\{I,X,Y,Z\}$ and the eigenvectors $\{|k\rangle\}$\ form the Bell-like
orthogonal basis
\begin{equation}
\left\{  \frac{|00\rangle+|11\rangle}{\sqrt{2}},\frac{|01\rangle+|10\rangle
}{\sqrt{2}},\frac{|10\rangle-|01\rangle}{\sqrt{2}},\frac{|11\rangle
-|00\rangle}{\sqrt{2}}\right\}  .
\end{equation}

Thus, one can compute the von Neumann entropy of the Choi matrix as the
Shannon entropy of the probability distribution $\mathbf{p}=\{p_{k}\}$, i.e.,
we may write%
\begin{equation}
S(\rho_{\mathcal{P}})=H(\mathbf{p})~, \label{SRC}%
\end{equation}
where $H(\mathbf{p}):=-\sum_{k}p_{k}\log_{2}p_{k}$. This means that the
(reverse) coherent information of a Pauli channel is%
\begin{equation}
I_{(R)C}(\mathcal{P})=1-H(\mathbf{p})~. \label{RCp}%
\end{equation}
In fact, a Pauli channel is unital, so that we can combine
Eqs.~(\ref{unitalRC}) and~(\ref{SRC}), to obtain Eq.~(\ref{RCp}).

Let us derive the entanglement flux $\Phi(\mathcal{P})$ of a Pauli channel.
One should compute the relative entropy of entanglement of its Choi matrix
$\rho_{\mathcal{P}}$ , i.e.,%
\begin{equation}
\Phi(\mathcal{P}):=E_{R}(\rho_{\mathcal{P}}):=\min_{\sigma\in\text{SEP}}%
S(\rho_{\mathcal{P}}||\sigma), \label{fluxP}%
\end{equation}
where, for two qubits, the set of separable states (SEP) coincides with the
set of states with positive partial transpose (PPT). The relative entropy at
the RHS\ of Eq.~(\ref{fluxP}) can be computed using the formula%
\begin{align}
S(\rho||\sigma)  &  =-S(\rho)-\mathrm{Tr}\left(  \rho\log\sigma\right)
\nonumber\\
&  =-S(\rho)-\sum_{i}\langle i|\rho|i\rangle\log s_{i}~,
\end{align}
where $|i\rangle$ ($s_{i}$) are the eigenstates (eigenvalues) of $\sigma$.

Finding the minimum in Eq.~(\ref{fluxP}) is hard in general. A very good
candidate is the following separable state
\begin{align}
\widetilde{\sigma}  &  :=\frac{1}{2}\sum_{u=0,1}|u\rangle\langle
u|\otimes\mathcal{P}(|u\rangle\langle u|)\nonumber\\
&  =\frac{p_{0}+p_{3}}{2}(|00\rangle\langle00|+|11\rangle\langle
11|)\nonumber\\
&  +\frac{p_{1}+p_{2}}{2}(|01\rangle\langle01|+|10\rangle\langle10|),
\end{align}
which is diagonal in the computational basis $\{|i\rangle\}=\{|00\rangle
,|01\rangle,|10\rangle,|11\rangle\}$, with eigenvalues $\{\tilde{s}_{i}\}$
given by
\begin{equation}
\left\{  \frac{p_{0}+p_{3}}{2},\frac{p_{1}+p_{2}}{2},\frac{p_{1}+p_{2}}%
{2},\frac{p_{0}+p_{3}}{2}\right\}  . \label{e2}%
\end{equation}

Thus, we compute the bound
\begin{equation}
\Phi(\mathcal{P})\leq S(\rho_{\mathcal{P}}||\widetilde{\sigma})=-S(\rho
_{\mathcal{P}})-\sum_{i}\langle i|\rho_{\mathcal{P}}|i\rangle\log\tilde{s}%
_{i}~,
\end{equation}
where $S(\rho_{\mathcal{P}})$ is given in Eq.~(\ref{SRC}). Using
Eq.~(\ref{Pcomp}), it is easy to check that
\[
-\sum_{i}\langle i|\rho_{\mathcal{P}}|i\rangle\log\tilde{s}_{i}=1+H_{2}%
(p_{1}+p_{2})~,
\]
where $H_{2}(p):=-p\log_{2}p-(1-p)\log_{2}(1-p)$ is the binary Shannon
entropy. Thus, for a Pauli channel with arbitrary distribution $\mathbf{p}%
=\{p_{k}\}$, we may write the bound
\begin{equation}
\Phi(\mathcal{P})\leq1+H_{2}(p_{1}+p_{2})-H(\mathbf{p})~. \label{PauliGEN}%
\end{equation}
This result can be specialized for the dephasing and depolarizing channels.

\subsubsection{Dephasing channel}

This is a Pauli channel with probability distribution $\mathbf{p}%
=\{p,0,0,1-p\}$, so that we have
\begin{equation}
\mathcal{P}_{\text{deph}}(\rho)=p\rho+(1-p)Z\rho Z~.
\end{equation}
It is easy to see that Eq.~(\ref{PauliGEN}) leads to%
\begin{equation}
\Phi(\mathcal{P}_{\text{deph}})\leq1-H_{2}(p)~.
\end{equation}
Note that this upper bound coincides with the lower bound give by the coherent
information $I_{C}(\mathcal{P}_{\text{deph}})=1-H_{2}(p)$ using Eq.~(\ref{RCp}%
). Thus, we get
\begin{equation}
Q_{2}(\mathcal{P}_{\text{deph}})=K(\mathcal{P}_{\text{deph}})=1-H_{2}(p)~,
\label{dephaRE}%
\end{equation}
which coincides with the unassisted quantum capacity of the channel
$Q(\mathcal{P})=1-H_{2}(p)$~\cite{degradable}.

\subsubsection{Depolarizing channel}

This is a Pauli channel with probability distribution
\begin{equation}
\mathbf{p}=\left\{  1-\frac{3p}{4},\frac{p}{4},\frac{p}{4},\frac{p}%
{4}\right\}  ,
\end{equation}
so that we have%
\begin{equation}
\mathcal{P}_{\text{depol}}(\rho)=\left(  1-\frac{3p}{4}\right)  \rho+\frac
{p}{4}(X\rho X+Y\rho Y+Z\rho Z)~.
\end{equation}
From Eq.~(\ref{PauliGEN}) we compute%
\begin{equation}
\Phi(\mathcal{P}_{\text{depol}})\leq1+H_{2}\left(  \frac{p}{2}\right)
-H_{2}\left(  \frac{3p}{4}\right)  -\frac{3p}{4}\log_{2}3~. \label{Bfirst}%
\end{equation}
This has to be compared with the following lower bound~\cite{BoundPRE}
\begin{equation}
Q(\mathcal{P}_{\text{depol}})\geq1-H_{2}\left(  \frac{3p}{4}\right)
-\frac{3p}{4}\log_{2}3~.
\end{equation}
Also note that the unassisted quantum capacity must satisfy
\[
Q(\mathcal{P}_{\text{depol}})\leq1-3p\text{~~~for }0\leq p\leq\frac{1}{3},
\]
and $Q(\mathcal{P}_{\text{depol}})=0$ otherwise.

We may improve the bound in Eq.~(\ref{Bfirst}) by resorting to the same
argument of Ref.~\cite{GEW}. Let us denote by $\mathcal{P}_{\text{depol}}^{p}
$ a depolarizing channel with probability $p$. Then, we may write the convex
combination
\begin{equation}
\mathcal{P}_{\text{depol}}^{p}=(1-\alpha)\mathcal{P}_{\text{depol}%
}^{\varepsilon}+\alpha\mathcal{P}_{\text{depol}}^{2/3},
\end{equation}
where $\alpha=(p-\epsilon)/(2/3-\epsilon)$ and $0\leq\epsilon\leq p\leq2/3$.
Here $\mathcal{P}_{\text{depol}}^{2/3}$ is entanglement-breaking, so that
$\Phi(\mathcal{P}_{\text{depol}}^{2/3})=0$. Then, using the convexity of the
entanglement flux, we may write $\Phi(\mathcal{P}_{\text{depol}}^{p}%
)\leq(1-\alpha)\Phi(\mathcal{P}_{\text{depol}}^{\varepsilon})$. Now, for any
$p$, we may consider an improved upper bound by minimizing over $\varepsilon$,
i.e.,%
\begin{equation}
\Phi(\mathcal{P}_{\text{depol}}^{p})\leq\min_{\epsilon}(1-\alpha
)\Phi(\mathcal{P}_{\text{depol}}^{\varepsilon})~.
\end{equation}

\subsection{Entanglement flux of the erasure channel}

This is not a Pauli channel. With some probability $p$, this channel replaces
an incoming qubit state $\rho$ with an erasure state $|2\rangle$, which is
orthogonal to it. In other words, we have the action
\begin{equation}
\mathcal{E}_{\text{erase}}(\rho)=(1-p)\rho+p|2\rangle\langle2|~.
\end{equation}
Its Choi matrix is given by
\begin{equation}
\rho_{\mathcal{E}_{\text{erase}}}=(1-p)|\Phi\rangle\langle\Phi|+\frac{p}%
{2}\left(  |02\rangle\langle02|+|12\rangle\langle12|\right)  ~.
\label{erasure1}%
\end{equation}
We construct the candidate separable state as before, i.e., we pick%
\begin{align}
\widetilde{\sigma}  &  :=\frac{1}{2}\sum_{u=0,1}|u\rangle\langle
u|\otimes\mathcal{E}_{\text{erase}}(|u\rangle\langle u|)\nonumber\\
&  =\frac{1-p}{2}\left(  |00\rangle\langle00|+|11\rangle\langle11|\right)
\nonumber\\
&  +\frac{p}{2}\left(  |02\rangle\langle02|+|12\rangle\langle12|\right)  ~,
\end{align}
which is diagonal in the computational basis.

Now by diagonalizing Eq.~(\ref{erasure1}), we compute the entropy
\begin{equation}
S(\rho_{\mathcal{E}_{\text{erase}}})=(1-p)\log(1-p)+p\log\left(  \frac{p}%
{2}\right)  ~. \label{Seerase}%
\end{equation}
Then, we derive
\begin{gather}
-\sum_{i}\langle i|\rho|i\rangle\log\tilde{s}_{i}=-\langle00|\rho
_{\mathcal{E}_{\text{erase}}}|00\rangle\log\left(  \frac{1-p}{2}\right)
\nonumber\\
-\langle11|\rho_{\mathcal{E}_{\text{erase}}}|11\rangle\log\left(  \frac
{1-p}{2}\right) \nonumber\\
-\langle02|\rho_{\mathcal{E}_{\text{erase}}}|02\rangle\log\left(  \frac{p}%
{2}\right)  -\langle12|\rho_{\mathcal{E}_{\text{erase}}}|12\rangle\log\left(
\frac{p}{2}\right)  ~. \label{ooerase}%
\end{gather}

Combining Eqs.~(\ref{Seerase}) and~(\ref{ooerase}), we derive the entanglement
flux (upper bound) of the erasure channel%
\begin{equation}
\Phi(\mathcal{E}_{\text{erase}})\leq S(\rho_{\mathcal{E}_{\text{erase}}%
}||\widetilde{\sigma})=1-p~.
\end{equation}
Note that the two-way quantum capacity of the erasure channel is already known
to be $Q_{2}(\mathcal{E}_{\text{erase}})=1-p$~\cite{ErasureChannel}. This
means that we have determined the secret-key capacity of this channel, since
\begin{equation}
Q_{2}(\mathcal{E}_{\text{erase}})\leq K(\mathcal{E}_{\text{erase}})\leq
\Phi(\mathcal{E}_{\text{erase}})~.
\end{equation}

\subsection{Entanglement flux of the generalized dephasing channel}

Consider the dephasing channel for a $d$ dimensional system. This channel has
Kraus representation~\cite{depha1,depha2}
\begin{equation}
\mathcal{P}_{d}(\rho)=\sum_{m=0}^{d-1}E_{m}\rho E_{m}^{\dag},~~E_{m}%
=\sqrt{P_{m}(p,d)}Z^{m},
\end{equation}
where
\begin{equation}
P_{m}(p,d)=\binom{d-1}{m}p^{m}(1-p)^{d-1-m}~,
\end{equation}
and $Z$ is defined in Eq.~(\ref{Pauli_DEF}).

Let us\ compute the Choi matrix $\rho_{\mathcal{P}_{d}}=(I\otimes
\mathcal{P}_{d})(\Psi_{d})$, where $|\Psi_{d}\rangle=d^{-1/2}\sum_{i=0}%
^{d-1}|ii\rangle$. We find
\[
\rho_{\mathcal{P}_{d}}=\sum_{m,i,j}^{d-1}\frac{P_{m}(p,d)}{d}\exp\left[
\frac{2i\pi}{d}(i-j)m\right]  |ii\rangle\langle jj|.
\]
By diagonalizing this density matrix, one finds $d$ non-zero eigenvalues
$\mathbf{P}:=\{P_{0}(p,d),\ldots,P_{d-1}(p,d)\}$, so that the Von Neumann
entropy of $\rho_{\mathcal{P}_{d}}$ is easily computed
\begin{equation}
S(\rho_{\mathcal{P}_{d}})=H(\mathbf{P})=-\sum_{m}^{d-1}P_{m}(p,d)\log_{2}%
P_{m}(p,d). \label{VonNe}%
\end{equation}

We now introduce the following optimal separable state (diagonal in the
computational basis)
\begin{align}
\widetilde{\sigma}  &  =\sum_{i=0}^{d-1}\frac{1}{d}|i\rangle\langle
i|\otimes\mathcal{P}_{d}(|i\rangle\langle i|)\nonumber\\
&  =\sum_{i,m=0}^{d-1}\frac{P_{m}(p,d)}{d}|ii\rangle\langle ii|=\sum
_{i=0}^{d-1}\frac{1}{d}|ii\rangle\langle ii|~,
\end{align}
where we have used $\sum_{m}P_{m}(p,d)=1$. Thus we can derive the second term
in the REE
\begin{align}
\mathrm{Tr}(\rho_{\mathcal{P}_{d}}\log\widetilde{\sigma})  &  =\sum
_{i=0}^{d-1}\langle ii|\rho_{\mathcal{P}_{d}}|ii\rangle\log_{2}\frac{1}%
{d}\nonumber\\
&  =\sum_{m}P_{m}(p,d)\log_{2}\frac{1}{d}\nonumber\\
&  =-\log_{2}d \label{pezzoTr}%
\end{align}

Combining Eqs.~(\ref{VonNe}) and~(\ref{pezzoTr}), we derive the following
result for the entanglement flux of the generalized dephasing channel
\begin{equation}
\Phi(\mathcal{P}_{d})\leq S(\rho_{\mathcal{P}_{d}}||\widetilde{\sigma}%
)=\log_{2}d-H(\mathbf{P}),
\end{equation}
which reduces to Eq.~(\ref{dephaRE}) for $d=2$ (qubits). Note that
$\Phi(\mathcal{P}_{d})\leq\log_{2}d-S(\rho_{\mathcal{P}_{d}})$ and, according
to Eq.~(\ref{unitalGEN}), the coherent information of this (unital)\ channel
is $I_{C}(\mathcal{P}_{d})=\log_{2}d-S(\rho_{\mathcal{P}_{d}})$. As a result,
lower and upper bounds coincide and we determine the two-way assisted
capacities
\begin{equation}
Q_{2}(\mathcal{P}_{d})=K(\mathcal{P}_{d})=\log_{2}d-H(\mathbf{P}).
\end{equation}


\begin{thebibliography}{99}
\bibitem {notaME}Note that maximally-entangled states are specific types of
private states~\cite{KD1}.
\bibitem {AmplitudeDamp}In such a channel, we have $\left\vert 0\right\rangle
\left\langle 0\right\vert \rightarrow\left\vert 0\right\rangle \left\langle
0\right\vert $ and $\left\vert 1\right\rangle \left\langle 1\right\vert
\rightarrow p\left\vert 1\right\rangle \left\langle 1\right\vert
+(1-p)\left\vert 0\right\rangle \left\langle 0\right\vert $ for some
probability $p$. Clearly, there is no unitary $U$ able to realize
$U\mathcal{E}_{\text{damp}}(\left\vert 0\right\rangle \left\langle
0\right\vert )U^{\dagger}=\mathcal{E}_{\text{damp}}(X\left\vert 0\right\rangle
\left\langle 0\right\vert X)$ for Pauli operator $X$.
\bibitem {NotaCOH}Furthermore, any other non-trace preserving local operation
which might be performed by Alice and Bob can be made completely coherent.
This is done by replacing measurements with suitable unitaries, ancillary
systems and including their outputs in the ensembles $\mathbf{a}$ and
$\mathbf{b}$.
\bibitem {Differenza}It is important to note the basic differences between
this procedure and the programmable quantum gate arrays~\cite{Processors}. In
our case, we are not considering a universal set of quantum operations applied
to an input system but a fixed quantum channel which suitably
\textquotedblleft commutes\textquotedblright\ with teleportation. Thus,
instead of post-selecting the identity outcome $\mathcal{U}_{k}=I$ (occurring
with probability $2^{-d}$ for a qudit and zero probability for a CV system),
we can consider all the outcomes of the Bell detection which are then
fed-forward in the final LOCC. Thus, our method departs from probabilistic
(i.e., non trace-preserving) programmable quantum processors and their
equivalent of port-based teleportation~\cite{Port-based} (requiring infinite
uses of the channel). In particular, the fact that our method provides an
overall trace-preserving LOCC is absolutely crucial for the simplification of
the adaptive protocols.
\end{thebibliography}
\end{document}